\documentclass[aps,apl,preprint,groupedaddress]{revtex4}

\usepackage{amsmath,amssymb}
\usepackage{graphicx}

\def\BibTeX{{\rm B\kern-.05em{\sc i\kern-.025em b}\kern-.08em
    T\kern-.1667em\lower.7ex\hbox{E}\kern-.125emX}}

\begin{document}

\title{Active split-ring metamaterial slabs for magnetic resonance imaging}
%
%
%
\author{Marcos A. Lopez$^1$}
\author{Jose M. Algarin$^1$}
\author{Manuel J. Freire$^1$}
\author{Volker C. Behr$^2$}
\author{Peter M. Jakob$^2$}
\author{Ricardo Marqu\'es$^1$}

\affiliation{$^1$Department of Electronics and Electromagnetism,
Faculty of Physics, University of Seville, Avda. Reina Mercedes
s/n, 41012 Sevilla, Spain } \email{freire@us.es}
\affiliation{$^2$Department of Experimental Physics 5
(Biophysics), Institute of Physics, University of
W$\ddot{u}$rzburg, Am Hubland, 97074 W$\ddot{u}$rzburg, Germany}

\date{\today}

\begin{abstract}
In this work, it is analyzed the ability of split-ring
metamaterial slabs with zero/high permeability to reject/confine
the radiofrequency magnetic field in magnetic resonance imaging
systems. Using an homogenization procedure, split-ring slabs have
been designed and fabricated to work in a 1.5T system. Active
elements consisting of pairs of crossed diodes are inserted in the
split-rings. With these elements, the permeability of the slabs
can be automatically switched between a unity value when
interacting with the strong excitation field of the transmitting
body coil, and zero or high values when interacting with the weak
field produced by protons in tissue. Experiments are shown for
different configurations where these slabs can help to locally
increase the signal-to-noise-ratio.
\end{abstract}


\maketitle

Application of metamaterials in magnetic resonance imaging (MRI)
has been previously explored in several works making use of
devices based on swiss-rolls \cite{Wiltshire}-\cite{Mathieu},
wires \cite{Radu} and capacitively-loaded split rings
\cite{Freire-APL-2008}-\cite{Freire-APL-2011}. Most of these works
have explored the sub-wavelength imaging ability of metamaterials
with negative permeability ($\mu$). In previous works of the
authors, metamaterials slabs with $\mu=-1$ has been fabricated and
tested in MRI systems to show the ability of these slabs to
increase the sensitivity of surface coils
\cite{Freire-APL-2008,Freire-JMR-2010} and to improve the field
localization of these coils, a fact that may find applications in
parallel MRI \cite{Freire-APL-2011}. Although metamaterials can be
engineered to tailor whatever value of permeability at the desired
frequency, little attention has been paid to permeability values
different from negative ones. In the present work, it is explored
the application in MRI of capacitively-loaded split-ring
metamaterials which show zero permeability ($\mu=0$) or high
permeability ($\mu\rightarrow\infty$) at the operating frequency.
$\mu=0$ and $\mu\rightarrow\infty$ slabs can reject (see Fig. 1.b)
and confine (see Fig. 1.c), respectively, the radiofrequency (RF)
magnetic field. These properties can help to locally increase the
signal-to-noise-ratio (SNR) of surface coils in certain
configurations which have been experimentally investigated in this
work. Typical MRI acquisition consists of the excitation of tissue
with a strong and uniform RF field generated by a transmitting
body coil, and then the detection of the weak field generated by
hydrogen nuclei in tissue by means of surface coils. The
split-ring device previously reported by the authors consisted of
a $\mu=-1$ slab \cite{Freire-APL-2008}-\cite{Freire-APL-2011},
which does not distort the uniform excitation field. However,
$\mu=0$ and $\mu\rightarrow\infty$ slabs can actually distort this
field. Therefore, it is necessary to implement these slabs as
active slabs which can be automatically switched to show
$\mu\simeq 1$ under the strong field of excitation and $\mu=0$ and
$\mu\rightarrow\infty$ under the weak field coming from tissue.
This can be accomplished in the practical implementation of the
slabs by inserting active elements in the split-rings that allow
to switch between different responses under strong or weak fields.
In particular, a pair of crossed diodes inserted in each
split-ring can help to switch off them under the strong excitation
field. Following the homogenization procedure previously reported
by some of the authors \cite{Baena-PRA-2008}, two split-ring slabs
of $6 \times 6\times 1$ unit cells with a periodicity of 15 mm
were designed to exhibit $\mu=0$ and $\mu\rightarrow\infty$ at the
frequency of 63.6 MHz. This frequency corresponds to the Larmor
frequency of 1.5T Siemens Avanto MRI system sited in the
Department of Experimental Physics 5 (Biophysics) of the
University of W$\ddot{u}$rzburg (Germany), where the experiments
reported in this work were done. The fabricated split-rings have
12 mm in diameter and 1.87 mm of strip width for the $\mu=0$ slab,
11.8 mm in diameter and 1.7 mm of strip width for the
$\mu\rightarrow\infty$ slab. Each split-ring in the array contains
a 470$\pm 1 \% $ pF non-magnetic capacitor (\emph{American
Technical Ceramics Corp.}, NY, USA) for resonance at a specific
frequency below 63.6 MHz, and a pair of crossed diodes
(\emph{Microsemi Corp.}, CA, USA) in parallel with the capacitor
(see Fig. 2) in order to switch off the slab in transmission.
Under the strong excitation field, the high electromotive force
induced in the rings makes the diodes to drive and then the
capacitors are short-circuited, so that the split-rings behave
like simple closed metallic rings. Following the homogenization
procedure \cite{Baena-PRA-2008}, the calculated permeability for
this system of simple metallic loops is $\mu=0.85$, a value which
is closed to the value of the permeability of air. Once the sample
is excited, the tissue reradiates a weak field which is unable to
drive the diodes, so that the rings behave like resonant circuits.
The frequency of resonance has been chosen so that from the
homogenization model \cite{Baena-PRA-2008} the system has $\mu=0$
and $\mu=\infty$ at the working frequency. Since the capacitance
of each split-ring is fixed, the frequency of resonance was fitted
by adjusting the dimensions of the rings. A small correction to
the value predicted by the homogenization model was necessary due
to the parasitic reactance of the diodes.

For the experiments, a 90 mm in diameter receive-only single loop
coil was used and a cylindrical bottle 16 cm in diameter, filled
with a H$_2$O solution doped with 5 g/l NaCl and 1.25 g/l
NiSO$_4$, was used as a load for the experiments. The loop was
tuned to 63.63 MHz and matched to $50\Omega$ in the presence of
the slabs and the phantom. It was actively decoupled by a tuned
trap circuit including a PIN diode in transmission. The active
decoupling for the loop was -25dB with and without metamaterial
slabs. All the experiments were performed in a 1.5 T whole body
scanner. In the $\mu=0$ experiment, the metamaterial slab is
perpendicular to the loop and it is positioned at one side of the
phantom (see Fig. 3), so that the magnetic flux is rejected by the
slab and then confined inside the phantom. This will increase the
signal coming from this region of the phantom. In the $\mu=\infty$
experiment, the metamaterial slab is placed parallel to the loop
in the opposite side of the phantom (see Fig. 4) in order to guide
the flux lines through the phantom.

SNR maps were calculated from a series of identical phantom
measurements \cite{Ohliger-MRM-2004} for both the $\mu=0$ and the $\mu=\infty$ slabs and
compared with the situation where the slabs were removed. In Fig.
3 top, the calculated SNR maps are shown for both the presence and
the absence of the $\mu=0$ slab and profiles are compared (see Fig. 3
bottom). In the side of the phantom where the $\mu=0$ slab is placed,
the signal increases (approx. 15\%). The calculated SNR maps for the
$\mu=\infty$ slab are shown in Fig. 4 top, and the corresponding profiles
in Fig. 4 bottom. Again, the signal presents an increment of
approximately 15\% with the presence of the slab.

This work demonstrates how split-ring metamaterial slabs designed
with specific permeability values can increase the SNR in
different configurations. The SNR gain in the demonstration was
moderate, but it could be improved with a smart design of the
configuration by suitably choosing both the coild and phantom
size. Moreover, although the SNR gain could not be comparable to
that provided by another loop coil positioned in the same as the
slab, the metamaterial slabs could be useful in limited channel
systems or as complement of an array. Some artifacts appear in the
phantom's surface due to the discrete nature of the split-rings,
but it can be easily removed by taking the salb 1 cm far from the
surface of the phantom.

This work was supported by the Spanish Ministerio de Ciencia e
Innovacion under project Consolider CSD2008-00066. The authors
want to thank the company NORAS MRI Products for the advice.

\newpage

\noindent {CAPTION TO FIGURES}

\bigskip

\noindent Figure 1. a) Magnetic field lines for a single coil. b)
Magnetic field lines with a $\mu=0$ slab perpendicular to the
coil. c) Magnetic field lines with a $\mu\rightarrow\infty$ slab
placed parallel to the coil.

\bigskip

\noindent Figure 2. Sketch of the constituent element and
photographs of the slabs. The slabs are $6\times 6\times 1$ unit
cells and the periodicity is 15 mm. The $\mu=0$ has been
fabricated with rings of 12 mm in diameter and 1.87 mm of strip
width, and the $\mu\rightarrow\infty$ has been fabricated with
11.8 mm in diameter and 1.7 mm of strip width. For both cases, 470
pF capacitors were used to tune the loops. A pair of crossed
diodes in parallel with the capacitors provides the active
response which is different for strong or weak fields.

\bigskip
\noindent Figure 3. SNR map for the experiment with the $\mu=0$
slab (up) in a axial plane. Profile of the SNR over the white line
in the map (down).

\bigskip

\noindent Figure 4. SNR map of the experiment with the
$\mu\rightarrow\infty$ slab(up) in a axial plane. Profile of the
SNR over the transversal axis (down).

\newpage

\begin{figure}[h]
\begin{center}
\includegraphics[width=8cm]{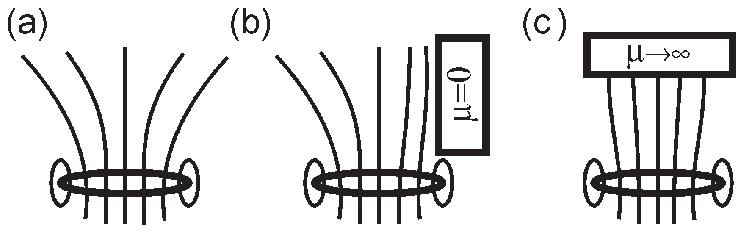}
\caption{}
\end{center}
\end{figure}
\begin{figure}[h]
\begin{center}
\includegraphics[width=8cm]{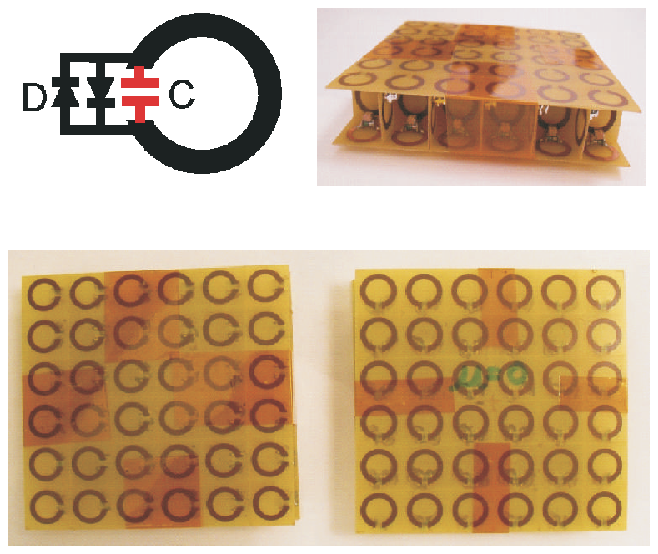}
\caption{}
\end{center}
\end{figure}
\begin{figure}[h]
\begin{center}
\includegraphics[width=8cm]{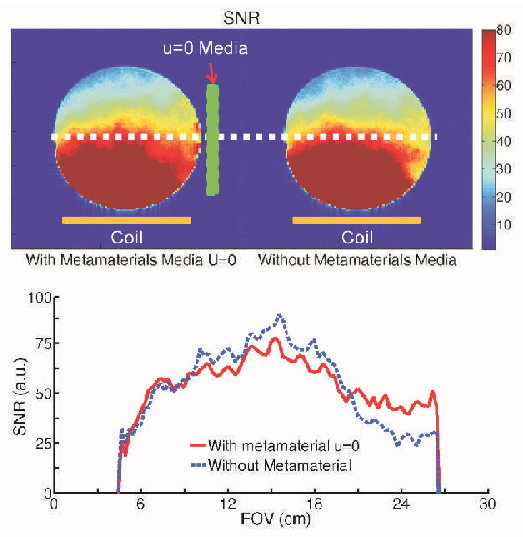}
\caption{}
\end{center}
\end{figure}
\begin{figure}[h]
\begin{center}
\includegraphics[width=8cm]{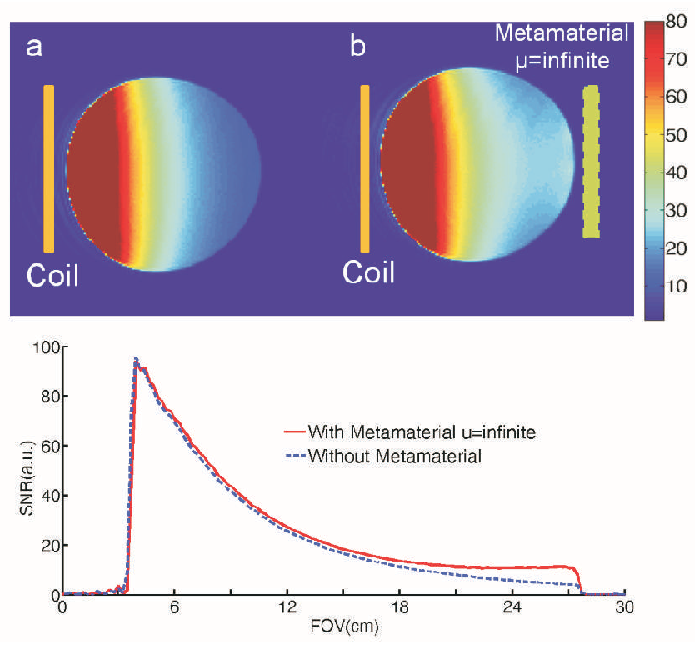}
\caption{}
\end{center}
\end{figure}

\end{document}